\documentclass[11pt, oneside]{article}   	
\usepackage{graphicx}				
\usepackage{amssymb}
\usepackage{amsmath}
\usepackage{amsfonts}
\usepackage{bm}

\title{Post-Model-Selection Statistical  Inference with Interrupted Time Series Designs: An Evaluation of an Assault Weapons Ban in California \thanks{
The California Department of Justice provided the background check data. Dorina Domi helped assemble information on potential interventions. To both I am grateful. Very useful feedback on an earlier version of this material was provided in response to an invited presentation to the Stanford/Berkeley Online Causal Inference Seminar in April of 2021. The discussant, John Donohue, was especially insightful. Thanks go to Susan Sorenson for educating me on firearm policy. A singular thanks to Arun Kumar Kuchibhotla for walking me through the multiple testing theory for post-model-selection inference.}}
\author{Richard Berk \\
University of Pennsylvania \\
berkr@sas.upenn.edu}

\begin{document}
\maketitle

\begin{abstract}
There have been many claims in the media and a bit of respectable research about the causes of variation in firearm sales. The challenges for causal inference can be quite daunting. This paper reports an analysis of daily handgun sales in California from 1996 through 2018 using an interrupted time series design and analysis. The design was introduced to social scientists in 1963 by Campbell and Stanley, analysis methods were proposed by Box and Tiao in 1975, and more recent treatments are easily found (Box et al., 2016). But this approach to causal inference can be badly overmatched by the data on handgun sales, especially when the causal effects are estimated. More important for this paper are fundamental oversights in the standard statistical methods employed. Test multiplicity problems are introduced by adaptive model selection built into recommended practice. The challenges are computational and conceptual. Some progress is made on both problems that arguably improves on past research, but the take-home message may be to reduce aspirations about what can be learned.
\end{abstract}

\section{Introduction}
The interrupted time series design was introduced to biomedical and social science researchers in the early 1960s (Campbell and Stanley, 1963). Its objective was to find a change in level or slope caused by an exogenous, discrete intervention for equally spaced longitudinal observations. An early application documented the impact of harsher penalties for speeding on yearly automobile crash fatalities (Campbell and Ross, 1968). Shortly thereafter, Box and Tiao (1975), building on the analysis of time series data, provided statistical procedures for the design that substantially furthered credible statistical and causal inference. Textbook treatments followed (Gottman and Glass, 1978; McDowell et al., 1980; Cryer and Chan, 2008: chapter 11; McDowell et al., 2019) along with related work in econometrics (Granger, 1980; Hamilton, 1994, Hendrey, 1995).

The strong quasi-experimental structure coupled with state-of-the-art methods of analysis provided a promising tool for the study of criminal justice interventions. Reforms that otherwise would be difficult to evaluate could be treated as social experiments (Campbell, 1969). Criminal justice researchers from a variety of academic disciplines recognized the potential (e.g., O'Carroll et al., 1991; Holmes et al., 1992; Webster et al., 2006; Vuji\'{c}, et al., 2016; Studdert et al., 2017; Levine and McNight, 2017; Liu and Wiebe, 2019).

Despite additional advances in methods for analyzing data from interrupted time series designs (Box et al.,  2016: Part 3), recent developments in statistics and econometrics have raised important concerns. They center on the way model specifications are inductively determined as an integral part of the data analysis, followed by statistical tests  confidence intervals undertaken with the same data. Recommended practice rests heavily on trial and error, guided by a variety of time series diagnostics. Over the past decade, powerful theoretical results have shown that such practices can be misguided (Leeb, and P\"{o}tscher, 2005; 2006; 2008). Pejorative characterizations include data snooping, fishing, data dredging, and p-hacking.\footnote
{
The issues are very general. They are not limited to time series model specification. Kuchibhotla and colleagues (2021:2) broadly address obtaining ``valid inference after data exploration'' (VIDE), with ``exploration'' to include a range of analysis activities such as examining correlation matrices, studying diagnostics of residuals, transforming response variables to stabilize their variance, and relying on automated procedures such as stepwise regression or the lasso.
}

Inductive model specification can introduce the well-known ``multiplicity problem.'' Multiple statistical tests and confidence intervals are undertaken as part of the same data analysis ignoring any inferential consequences. In particular, a nominal critical value for $\alpha$, such as the .05 level, is no longer the actual critical value. The actual critical level is larger, often very much larger. One consequence is a greater risk for Type I errors; null hypotheses are more easily rejected when they are true. The result is false discoveries that contribute to current reproducibility controversies in the social and biomedical sciences (Ioannidis, 2005; 2012; Scholler, 2014).   

There are excellent monographs and textbooks on different kinds of multiplicity problems and a wide range of proposed remedies (Bretz et al., 2011; Maxwell et al., 2017: chapter 5; Shirairshi et al., 2019). However, existing treatments do not extend multiplicity concerns to model specification. Indeed, trial and error specification procedures are commonly recommended in respected sources for a variety of popular data analysis methods as if multiple testing were irrelevant (Weisberg, 2014: Chapters 9-10).

This paper addresses the multiplicity problem for interrupted time series analysis. The approach taken is an extension of work in statistics on post-model-selection statistical inference that has emphasized cross-sectional data (Berk, 2013; Tibshirani et al., 2016; Lee et al., 2016; Kuchibhotla et al., 2021). In the pages ahead, solutions for interrupted time series data are offered, although for formal conceptions of causal inference, important interpretive complications remain. The issues are illustrated with an interrupted time series analysis of the possible impact of an assault weapons ban on handguns sales in California. 

To set the stage, Section 2 provides a brief overview of analysis procedures for the interrupted time series design. With the problem to be addressed introduced, Section 3 summarizes generic solutions for multiplicity difficulties and remedies that can be appropriate for interrupted time series data. In Section 4, the focus turns to post-model-selection inference for interrupted time series data. Section 5 introduces the data that are analyzed in Section 6. Conclusions are offered in Section 7.

\section{A Brief Overview of Analysis Procedures for an Interrupted Time Series Design}

The power of interrupted time series analyses derives from a substantial number of equally spaced, temporal observations before and after an intervention of interest. Potential confounders that change gradually can be statistically controlled by adjusting for apparent trends as part of the model specification, even if the sources of such trends are unknown. The major threat to internal validity is other abrupt events that overlap in time with the principal intervention. If they are known in advance or easily recognized, they often can be controlled as part of the model specification. However, that can turn a model specification problem into estimation problem insofar as the principal intervention and abrupt confounders are substantially correlated. Such concerns typically favor very spare models. The data analysis discussed later proceeds in this fashion.

The standard exposition of time series analysis, including the role of interventions, is found in a justly famous book by G.E.P. Box and several colleagues that has gone through five editions (Box et al., 2016). Beneath the details, an analysis of data from an interrupted time series design is essentially a regression analysis with dependent disturbances, much as in the spirit of parametric, generalized least squares. Many of the usual statistical intuitions carry over, but there can be important differences in the particulars. For those interested in these specifics, they are summarized in the next four pages. Some readers may prefer to skip to the next section.

All frequentist statistical inference requires a stochastic process responsible for generating the data. Sometimes the data are generated by a researcher using random assignment or probability sampling. The former is standard in randomized control trials. The latter is standard in sample surveys. However, there are a wide variety of settings in which a valid data generation mechanism must be postulated.

The analysis of data from an interrupted time series design begins with an assumed underlying data generation process derived from a limitless, linear combination of weighted, white noise perturbations going back forever in time. More formally,
\begin{equation}
Y_{t} = e_{t} + \psi_{1}e_{t-1} + \psi_{2}e_{t-2} \dots,
\end{equation}
where for times $(t), (t-1), (t-2), \dots, (t-\infty)$,  the $Y_{(.)}$ are a time series of values, the $e_{(.)}$ are white noise perturbations, and the $\psi_{(.)}$ are weights that can differ over perturbations.\footnote
{
The white noise perturbations are generated i.i.d; they are independently realized from a single probability distribution with a mean of 0.0. Sometimes the distribution is assumed to be normal. For large samples, the normality assumption is, in practice, not important. 
}

In words, an observed time series such as automobile crash fatalities each year is assumed to be generated by a linear combination of white noise ``shocks'' with all but the most recent weighted by a corresponding constant. The signs and relative sizes of the weights determine the dependence in $Y_{t}$. For example, if the weights close in time to $Y_{t}$ are large, values of the time series more proximate in time will be more strongly related than values of the time series that are less proximate in time. 

Constraints are placed on the weights. In particular, it is common to require
that
\begin{equation}
\sum_{t=1}^{\infty} \psi^{2}_{t} < \infty \hspace{.5in} {\rm and} \hspace{.5in} E(Y_{t}) = 0.
\end{equation}
The sum of the squared weights is bounded, and the expectation of the time series is 0.0. In combination, the two define a common form of ``weak'' stationarity; the time series is centered on a mean of zero with $Cov(Y_t, Y_{t-k}) = \sigma^2_e \sum_{i=0}^\infty \psi_i \psi_{i+k}$.\footnote
{
The centering at 0.0 is not formally necessary. Centering on some other expected value will suffice. The time series then varies around some other constant than 0.0.
}
This means that the time series has a stable level, and the covariance between observations depends only on the the number of time periods between them (i.e., $k$). Neither the expected level nor the covariance change with respect to any arbitrary displacement of time. Stationarity is essential for the full array of time series analysis statistical tools to perform properly.   

The white noise perturbations are unobservable, and the researcher only gets to see $y_{t}$, which are realizations $Y_{t}$. Consequently, all subsequent models are \textit{approximations} of the true data generating process (Box et al., 2016: Section 1.3; Chapter 6).  Because the time series can be altered by one or more interventions, accommodations are made for fixed exogenous variables. But, no claims are made the any specification is correct by the usual regression definitions. Moreover, the approximation does not need to be unique. Several different approximations can suffice.

 The approximations for an interrupted time series take the general form:
\begin{equation}
y_{t} = f(\kappa,\zeta, t) + N_{t},
\end{equation}
where $f(\kappa,\zeta, t)$ is the mean function $m_{t}$ with $\kappa$ denoting parameters such as regression coefficients, $\zeta$ denoting interventions each represented as a step function or pulse, and $t$ as an index for time. 

\begin{figure}
\begin{center}
\includegraphics[scale=.20]{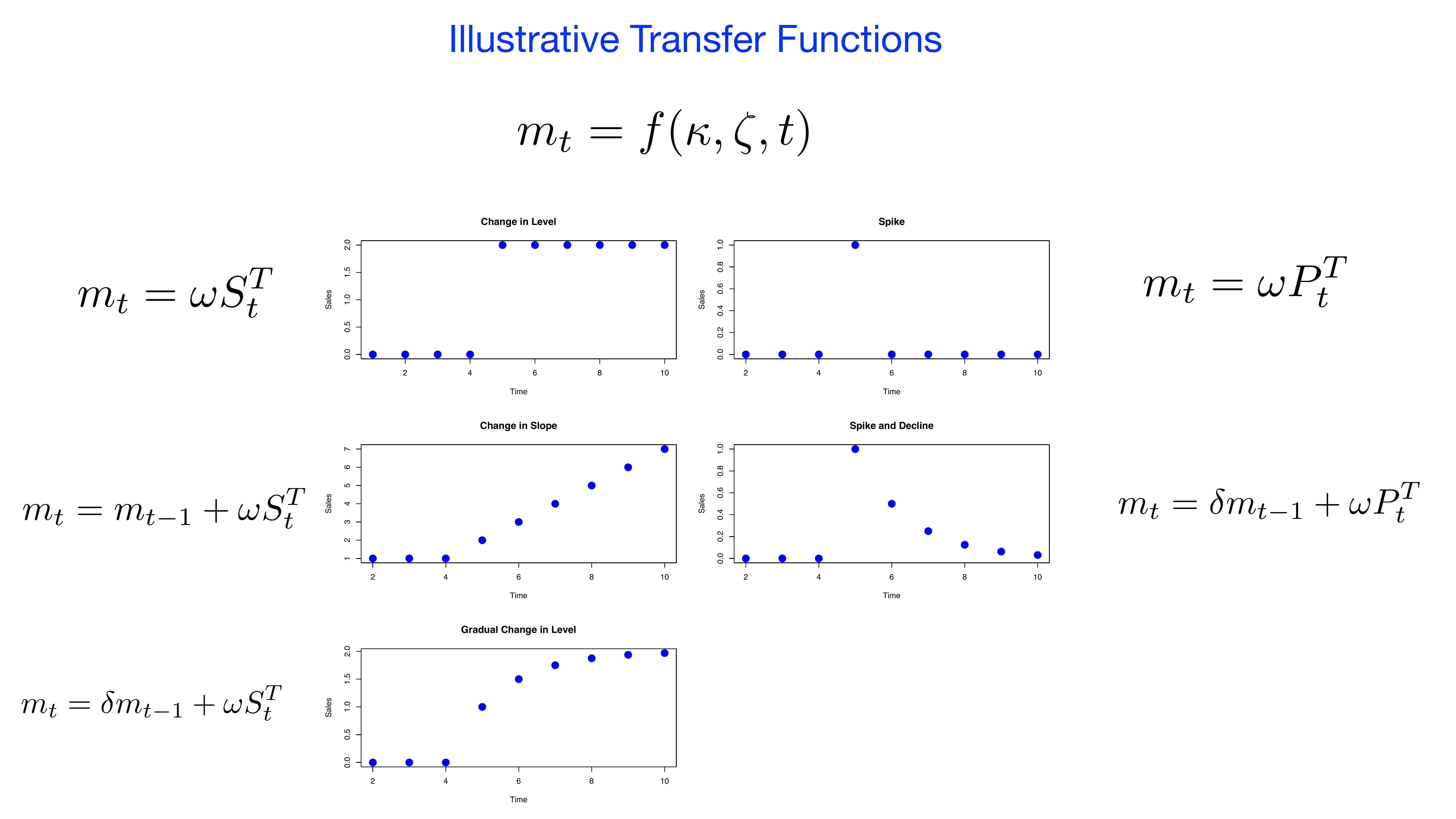}
\caption{Some Common Means Functions for Step Functions $(S_{ t})$ and Pulses $(P_{t})$}
\label{fig:transfer}
\end{center}
\end{figure}

The mean function $m_{t}$ can take a variety of forms, often called ``transfer functions" in the time series literature. Figure~\ref{fig:transfer} shows some common examples. An intervention is coded as a 1 or 0 for step function $S_{t}$ or as a 1 or 0 for pulse function $P_{t}$. A pulse has only a single value of 1 when the intervention occurs with all other values equal to 0.0. The size of the immediate impact on $y_{t}$ is determined by the value of $\omega$. The time path of that effect is determined by the value of $\delta$. In Figure~\ref{fig:transfer}, $\delta \le |1.0|$. Five different time paths are shown along with their corresponding mean functions. Allowance can be made for more than one intervention, but even with a large number of observations, multiple interventions can lead to substantial instability.

There can be, however, an important role of additional regressors in the mean function, whether seen as interventions or not. Their job is to adjust for possible confounding even when they are not part of the subject-matter account. Much as in the spirit of RCTs, the goal is not to construct a general model of some phenomenon, but to estimate the impact of one or more ``treatments.''\footnote
{
In some settings, the interrupted time series design is seen as a special case of single subject designs, which do not yield general models, but an estimate of an intervention's causal effect.
}
 
$N_{t}$ is a function for the residuals whose main purpose is to ``mop up'' any remaining dependences that would otherwise undermine statistical tests and confidence intervals. Commonly, it has an ARMA form $\phi(B)N_{t} = \theta(B)a_{t}$, with $B$ a polynomial backward shift operator and $a_{t}$ a new white noise disturbance.\footnote
{
$(1-B- B^2)y_{t} = y_t - y_{t-1} - y_{t-2}$ is an example of a polynomial backward shift operator that produces $y_{t}$ for three different time lags. 
}
An ARMA formulation allows for lagged values of the response variable $y_{t}$ and lagged values of the new white noise perturbations $a_{t}$. The ``AR'' component denotes an autoregressive process. The ``MA''  denotes a moving average process. Rarely is either given an explanatory role. The heavy causal inference lifting is done by the mean function $m_{t}$.

The ARMA structure is often extended in two ways.\footnote
{
The notation becomes rather involved, and will not be provided here. A very accessible treatment can be found in the textbook by Cryer and Chan (2008: Chapter 10).
}
First, to address deterministic trends that would violate the required stationarity, the time series data can be differenced. A first difference, represented as $\Delta(y_{t}) = y_{t} - y_{t-1}$, removes a linear trend. A second difference, represented as $\Delta^{2}(y_{t}) = [(y_{t} - y_{t-1}) - (y_{t-1} - y_{t-2})] = y_{t} - 2(y_{t-1}) + y_{t-2}$, removes a quadratic trend. Higher order differencing removes higher order trends. The goal is to restore stationarity violated by a change in level. One now has an ARIMA model for the disturbances instead of an ARMA model.\footnote
{
The ``I'' in ARIMA denotes the possible use of differencing and stands for integration. This is misleading because ARIMA models are based on difference equations for which the analog to integration is summing, which reverses differencing. 
}

Second, to address ``seasonal'' dependence in the residuals, a seasonal ARIMA formulation is often employed in a multiplicative form. The seasonal formulation captures dependence at lags with gaps. With a daily time series, for instance, a seasonal ARIMA structure can consider dependence at lags of 7 days, such that there is a temporal association between all Sundays, all Mondays, and so on, for each day of the week. In other words, all Sundays are more alike than all Sundays and all Mondays. The dependence can be removed by including seasonal differencing, seasonal AR terms, and/or seasonal MA terms. The multiplicative specification provides for interaction effects. For example, the dependence between $a_{t}$ and $a_{t-1}$ can differ depending on day of the week. There may be a stronger dependence between Mondays and Tuesdays than between Sundays and Mondays. 

Rarely is the approximation's specification known before the data are examined. There is a range of empirical specification aids routinely used. Graphical representations combined with statistical tests can be applied to the residuals of the mean function in search of acceptable forms for $N_{t}$. Many different prospective functions for $m_{t}$ and $N_{t}$ can be estimated, and the residuals examined. The goal is to find at least one specification of $m_{t}$ and $N_{t}$ for which one cannot reject the null hypothesis that $a_{t}$ is a white noise process. Any specification meeting this requirement is an acceptable approximation. There can be several such approximations; the model selected need not be the only  satisfactory one.

Some specifications require an estimator for nonlinear regression relationships. There are effective maximum likelihood and conditional least squares estimators that can work well.\footnote
{
The least squares solution is called conditional because one needs information before the earliest value of $y_{t}$ to implement estimation (e.g., $y_{t}$ depends on $y_{t-1}$). Such values are determined as form of imputation that often is quite simple. For example, an unobserved $y_{t-1}$ may be imputed using the empirical mean of the time series. Imputations become more complicated with longer lags common in seasonal ARIMA formulations. The least squares estimates are conditional on such imputations. Fortunately, any biases that result become less important the more time series observations available going forward.
}
Unfortunately, the lagged relationships often produce strong associations between the estimated parameters, leading to very unstable results and sometimes convergence failures. There also can be inadvertent parameter redundancy (Box et al., 2016: Section 7.3.5) Therefore, model specification must be highly selective in the predictors chosen. The usual intuitions for cross-sectional data are insufficiently precautionary. These issues are revisited later when the application is undertaken.

To summarize, model specification for an intervention analysis in time series traditions is a cherry picking exercise. Even if several approximations ultimately are selected, there is a price to paid for searching over many prospective models and picking a few (or one) for causal interpretation. It does not matter whether statistical tests were actually used to inform the searching. They are implicated anyway; \textit{all} cherry picking becomes a multiple testing problem. To consider these issues more specifically, the next section provides a brief exposition of multiple testing problems and solutions.

\section{A Brief Overview on the Multiplicity Problem}

There are many ways to formulate the multiple testing problem. Emphasized here
is an approach that dovetails well with the analysis of interrupted time series data. A simple, widely used, exposition will suffice. Readers already familiar with the Bonferroi and max-t corrections for multiplicity may prefer to skip to Section 4 where post-model-selection inference is discussed as a multiple testing problem.  

For a given dataset, there is more than a single null hypothesis to be tested, often many. Each test is evaluated separately at some nominal, critical value $\alpha$, often set in advance to .05 or .01. A null hypothesis is rejected if, under the null hypothesis, the test statistic, such as a t-statistic, would occur with a probability less than or equal to $\alpha.$ The same decision rule is used for each test. Concerns center on incorrect rejections of null hypotheses that are true, sometimes called ``false positives'' or Type I error. With $\alpha=.05$, say, Type I error is tolerated for a single test only 5\% of the time. The weight of the empirical evidence must be strong before the null hypothesis is rejected. This protection is eroded when more than one test is undertaken.

For $\alpha=.05$, let $m$ equal the number of statistical tests undertaken. For $m=2$, the actual critical value is $ 1-(1-\alpha)^m = .0975,$ which is substantially larger than .05, and it increases rapidly with number of tests. With 10 tests, the operational critical value is approximately .40, nearly an order of magnitude larger than .05. One has ``fake'' power leading to a much greater chance of Type I errors. Fake power can be seductive, and is one of the explanations offered for the ``reproducibility crisis'' in science, especially in the biomedical and social sciences (Ioannidis, 2005; 2012; Schooler, 2014).

Fake statistical power can materialize in more complicated ways that depend on the null hypotheses tested and the manner in which the data are generated (e.g., regression coefficient contrasts for a linear regression model). It follows that the proposed corrections are many and varied, often with  power concerns.\footnote
{
A Type II error is not rejecting a false null hypothesis. If the probability of a Type II error is denoted by $\beta$ (e.g., .2=0), power is $1-\beta$. Just as for Type I error, there are several different definitions are Type II error in a multiple testing context (Bretz et al., 2011) that lead to different definitions of power. 
}
In this paper, we focus on problems that can arise for parametric regression models, especially as they are applied to time series data. But they apply to nonparametric models as well.

\subsection{Two Error Rate Definitions}

Table~\ref{tab:errors}, taken from Bretz and colleagues (2001: page 12), can be used to help  formulate the error rate definitions. For a particular study, there are $m$ statistical tests overall. For example, a study of sentencing might begin with tests for disparities in median sentence lengths between different racial and ethnic groups. The null hypothesis is that the difference in medians between any two racial or ethnic groups is 0.0. If there are 5 such groups, there are 10 statistical tests. In Table~\ref{tab:errors}, therefore, $m=10$. 

\begin{table}[htp]
\caption{Outcomes from Multiple Statistical Tests}
\begin{center}
\begin{tabular}{|c|l|l|c|}
\hline \hline
Hypotheses & \hspace{.1in} Not Rejected & \hspace{.4in}Rejected & Total \\
\hline\
True            &  U (Correct Result)  &  V (Type I error)       & $m_{0}$ \\
 False          &  T  (Type II error)   &  S  (Correct Result)    & $m - m_{0}$ \\
\hline\
Total           &   \hspace{.65in} W   &  \hspace{.65in} R       & \hspace{.01in} $m$ \\
\hline \hline 
\end{tabular}
\end{center}
\label{tab:errors}
\end{table}

The observable counts $W$ and $R$ are the number of times the null hypothesis is not rejected and the number of times the null hypothesis is rejected, respectively. Because one does not know in advance the actual number of times the null hypothesis is true (i.e., $m_{0})$ or the number of times the null hypothesis is false (i.e., $m-m_{0})$, the cell counts $U$ through $S$ also are unknown. For instance, $U$ in this case could be any integer from 0 to 10. 

For a given a value of $m$ and a given critical value $\alpha$, many different definitions of the multiplicity error have been proposed (Bretz et al., 2011: Section 2.1). Immediately below are arguably the two most common definitions of multiplicity error.
\begin{itemize}
\item
\emph{Familywise Error Rate}: FWER=$\mathbb{P}(V>0)$ --- the probability in a set of $m$ hypothesis tests that there will be at least one Type I error. 
\item
\emph{False Discovery Rate}: FDR=$\mathbb{E}(V/R | R > 0)\mathbb{P}(R>0)$ --- the expected proportion of incorrectly rejected null hypotheses for all $R$ rejected null hypotheses (i.e., the expected proportion of rejected null hypotheses that are Type I errors). 
\end{itemize}
The FWER is equal to or larger than the FDR, because the FWER addresses the probability of at least one false discovery whereas the FDR addresses the expected proportion of false discoveries. No matter which of the two definitions is used, the goal is to employ statistical tests that provide a guarantee that the FWER or the FRD are controlled at the intended threshold. For example, the FWER intended threshold $\mathbb{P}(V>0)$ might be a conventional $\alpha=.05$. Special techniques that adjust for multiple tests are needed to guarantee that that $\mathbb{P}(V>0)$ is not greater than .05. Similar reasoning applies to the FDR. 

Control can be of two types. The control is weak if it applies only when all of the null hypotheses are true. This is sometimes called the ``global null hypothesis.'' Control is strong if it applies to all possible configurations of true and false null hypotheses. Strong control is desirable because although in practice a researcher will not know how many null hypotheses actually are false, it is unlikely that the research would have been undertaken unless some were thought to be. Controlling strongly for FWER with a given value of $\alpha$ automatically controls strongly for the FDR using that same $\alpha$ (Bretz et al., 2011: 14). 

\subsection{The Bonferroni Correction}

It is only modest exaggeration to see the Bonferroni correction as the ``mother'' of all multiplicity corrections. For, $m$ independent tests of true null hypotheses $H_1 , H_2 \dots H_m$, and rejection probabilities $\mathbb{P}(R_{1}),  \mathbb{P}(R_{2}) \dots, \mathbb{P}(R_{m})$ each set to $\alpha$, a Bonferroni adjusted critical value $\alpha^{\ast}$ is (Rice, 1995, 11.4.8),
\begin{equation}
\begin{split}
\alpha^{\ast} & = \mathbb{P}(R_1 \hspace{.05in} or  \hspace{.05in} R_2  \hspace{.05in} or \dots R_m) \\
& \le \mathbb{P}(R_1) + \mathbb{P}(R_2) + \dots + \mathbb{P}(R_{m}) \\
& = m\alpha.
\end{split}
\label{eq:bonferroni1}
\end{equation} 

The first line in Eq.~\ref{eq:bonferroni1} defines the actual p-value $\alpha^{\ast}$ that is being used because of $m$ statistical tests. It is the probability that \textit{any} true null hypothesis will be rejected. By the second line Eq.~\ref{eq:bonferroni1}, this must be equal to or less than the sum of the probabilities of rejecting the null hypothesis at a given value of $\alpha$, which in line 3 equals $m\alpha$.  In other words, a researcher thinks that null hypotheses are being rejected at $\alpha$ when they are really being rejected at $m\alpha$.

The Bonferroni correction is easily implemented. A new critical level $\alpha/m$ can be employed instead of $\alpha$. For example  if $m = 8$ and $\alpha =.01$, the adjusted probability threshold for ``statistical significance'' is $.001$. More formally, let $i$ denote a particular computed p-value $p_{i}.$ A computed p-value $p_{i}$ must be equal to or less than $.001$ for the null hypothesis to be rejected. Equivalently, $\alpha =.01$ can be retained, and each $p_{i}$ adjusted upward such that $q_{i}= mp_{i}$. For $p_{i} =.03$, say, $q_{i}=.24$.

The Bonferroni correction imposes strong control over the familywise error rate even if some of the null hypotheses are truly false. Bretz and colleagues (2011: 31) show that
\begin{equation}
\begin{split}
\mathbb{P}(V > 0) & = \mathbb{P} \left [ \bigcup_{M_{0}} (q_{i} \le \alpha) \right ]  \\
& \le \mathbb{P}  \sum_{M_{0}} (q_{i} \le \alpha)  \\
& \le m_{0}\frac{\alpha}{m} \le \alpha.
\end{split}
\label{eq:bonferroni2}
\end{equation} 
The first line in Eq.~\ref{eq:bonferroni2}, for the union of properly rejected tests, is the relationship one is seeking. However, it cannot be directly computed because the potential dependence between the tests is unknown. The next two lines show relationships that, when there is dependence, the Bonferroni  correction is conservative. It overcorrects such that power is reduced unnecessarily and confidence intervals are too wide. Also, the guarantee is asymptotic unless the relevant variables are normally distributed. This is no different from the properties of a conventional t-test, although the asymptotic requirement for non-normal data is too often ignored.  

The loss of power because of dependent tests is usually the primary motivation for a host of Bonferroni variants that can recover at least some the losses. Each has its idiosyncrasies, and performance will usually depend on the nature of any dependence between the test statistics. For example, the Holm procedure (Holm, 1979)``uniformly improves on the Bonferroni approach'' (Bretz et al., 2011: 32) by using a step by step procedure.  All p-values are ordered from small to large. The data analyst then proceeds in that same order, one p-value at a time, applying a variant of the usual Bonferroni correction that takes the number previous steps into account. Looking at Eq.~\ref{eq:bonferroni1}, the value of m is reduced by 1 at each step so that there are smaller Bonferroni corrections as the stepping from 1 to $m$ p-values proceeds. However, the reliance on the Eq.~\ref{eq:bonferroni1} means that the Holm procedure still can be conservative because the upper bound guarantee of  $\mathbb{P}(R_1) + \mathbb{P}(R_2) + \dots + \mathbb{P}(R_m) $ remains.

Another way to improve power is to abandon the FWER in favor of the FDR and use a step by step approach in the spirit of the Holm procedure (Benjamini and Hochberg, 1995). Independent test statistics formally are required, but more recent work  has proved that satisfactory results can be obtained if there are positive associations between the test statistics for the true null hypotheses (Benjamini and Yekutieli, 2001).\footnote
{
One can sometimes make progress with superficially small changes in the error rate definition. For example, the positive false discovery rate pFDR = $\mathbb{E}[\frac{V}{R} | R > 0].$ This addresses difficulties that can result if no null hypotheses are rejected (Bretz et al., 2011: 13-14), which can easily happen when power is weak. 
}
Once again, however, there can be tradeoffs. The Benjamini and Hochberg approach offers greater power than the classical Bonferroni approach, but at the price of more Type I errors (Bretz et al., 2011:14). 

\subsection{A Max-t Correction} 

A major disadvantage of Bonferroni correction and it variants is the reliance on an upper bound guarantee. Greater power can be obtained from direct estimates of the test statistic distribution itself (Romano and Wolk, 2005; 2017). For this, consider the ``max-t'' correction (Bretz et al., 2011: 21). Using the FWER definition of error and resampling procedures, one can bypass the Bonferroni upper bound while automatically adjusting for multiple, dependent t-statistics. If the distribution of the t-statistics is well estimated, corrected p-values follow. Building on the distribution of t-statistics is a powerful approach that can counter even very aggressive forms of p-hacking.

The formulation is relatively straightforward. Suppose one is comparing the proportion of probation sentences given by different judges in a particular jurisdiction. Suppose there are 8 such judges, leading to twenty-eight pairwise comparisons. The global null hypothesis is that all of the proportion differences are equal to 0.0; all judges have the same probability of giving a sentence of probation. The value of $\alpha$ set at .05. But, the operational p-value is .76. There actually is a 76\% chance that a ``statistically significant" difference in proportions will be found for at least one comparison even if the global null hypothesis is true.\footnote
{
In practice, more would need to be known about how the data were generated to make such a claim. For example, a simple random sample of judges from the jurisdiction could justify such a claim.
}
The usual Bonferroni correction would require a probability of .002, rather than .05 to reject the global null hypothesis, but the .002 level is conservative insofar as the 28 tests are not independent.

Alternatively, one can construct, say, 100 bootstrap samples of the data.\footnote
{
If the data were originally generated as a simple random sample of all judges in a jurisdiction, the ``plain vanilla'' nonparametric bootstrap would provide valid results, although the bias-corrected accelerated bootstrap, might perform somewhat better (Efron and Tibshirani, 1993: chapter 14). For dependent data, such as a time series, one must use a dependence-aware resampling method for which there are several options. One such option is employed later.
}
 For each, 28 t-tests are computed, and the largest of the 28 stored. There would be 100 such maximum values of the t-statistic from which one has a bootstrap estimate of the sampling distribution of the largest t-statistics over realizations of the data that automatically and properly allows for any dependence between tests. The .025 quantile and the .975 quantile then provide the two thresholds for rejection regions, assuming a two tailed test. To reject the global null hypothesis, at least of one of the original 28 t-statistics must fall in either of the rejection regions. Max-t procedures will have more power than the Bonferroni correction for a given value of $\alpha,$ with no increase type I error. These ideas form the foundation for how multiple testing challenges for an interrupted time series analysis are addressed later.
 
 \section{Post-Model-Selection Statistical Inference} 
 
Within its frequentist framework, classical statistical inference requires that each null hypothesis is specified before the data analysis begins. Because statistical tests will vary depending on the model, canonical statistical inference for modeling requires that the model also is specified before the data are examined. If a model is developed as part of the data analysis, estimation, statistical tests, and confidence intervals can be badly compromised.\footnote
{
It is a bit like a poker game in which no player is allowed to look at the cards any opponent is holding.  Bets are placed with no direct knowledge of any opponent's hand. If a player is caught peeking, even briefly, the game is immediately halted. Cheating has occurred. 
}

 \begin{figure}
\begin{center}
\includegraphics[width=3.5in]{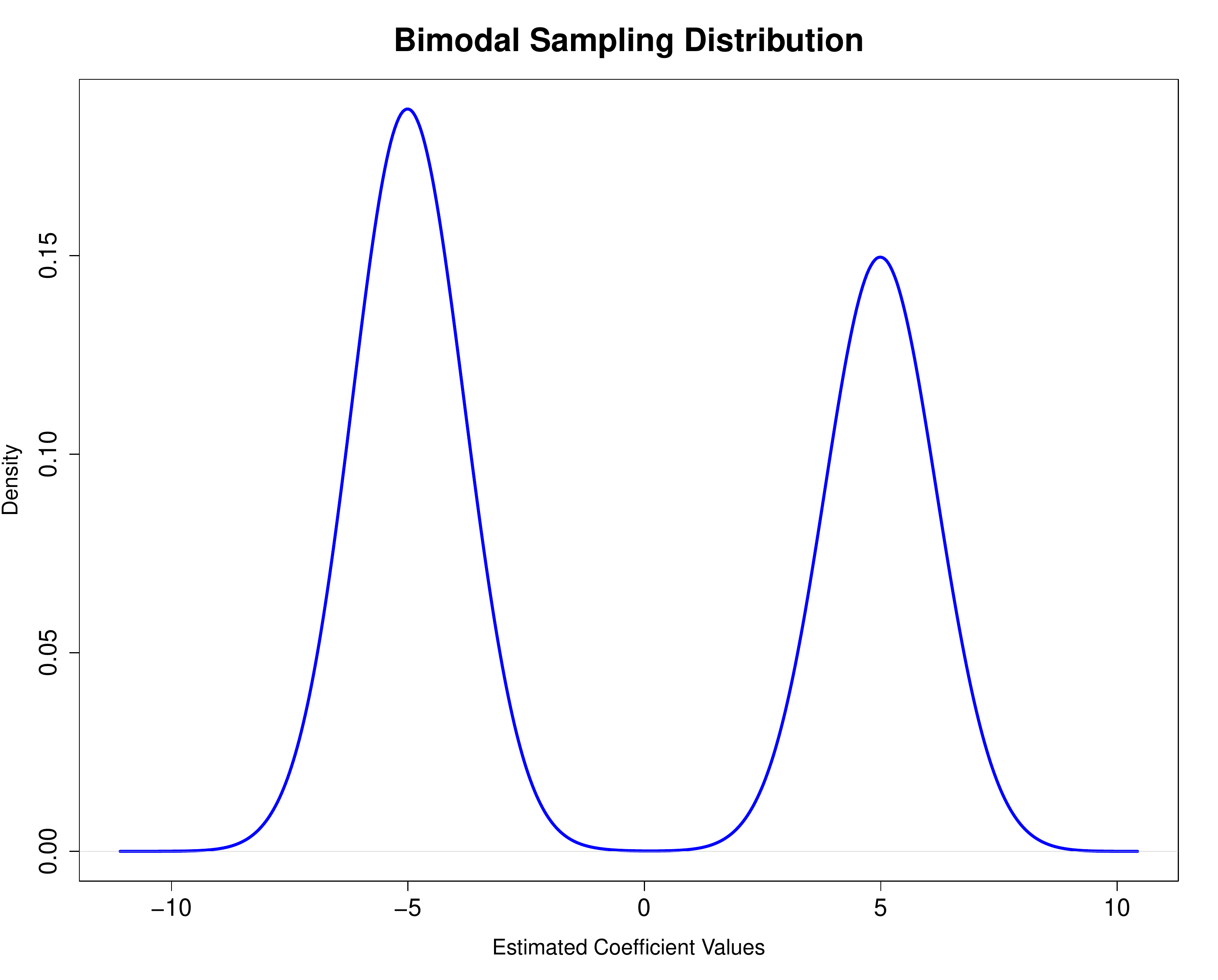}
\caption{A Bimodal Sampling Distribution for $\hat{\beta}_{1}$ Depending on the Model Selected}
\label{fig:bimodal}
\end{center}
\end{figure} 

Suppose, for example, a linear regression model needs to be specified. There are five predictors $X_{1}, X_{2} ,\dots, X_{5}$. $\beta_{1}$, the regression coefficient for $X_{1}$, is one parameter of interest. A researcher applies a model selection procedure such as forward stepwise regression, although less formal means could be used instead. Figure~\ref{fig:bimodal} shows some important consequences for the  $\hat{\beta}_{1}$ sampling distribution for over i.i.d realizations of the data. Because $X_{1}$ is only chosen for a selected model when  $\hat{\beta}_{1}$ is substantially different from 0.0, a bimodal sampling distribution results. The required normal sampling distribution will not materialize, even asymptotically. All classical statistical inference loses its desirable properties. This presents a major challenge for valid post-model-selection inference.

There is no argument about the consequences for post-model-selection inference. Formal proofs have existed for over 15 years that go well beyond baldly inappropriate research practices such p-hacking, data dredging, and data snooping (Leeb, and P\"{o}tscher, 2005; 2006; 2008). The risks are no less for inductive statistical procedures such as all subsets regression, recursive partitioning, model revisions after examining diagnostic plots or specification tests, and analysis procedures such as smoothing whose tuning parameter values commonly are determined as part of the fitting process (Berk et al., 2009). 

Post-model-selection inference can be productively reformulated as simultaneous inference (Kuchibhotla et al., 2021). It begins with a universe of models that could be realized, specified by the researcher in advance of the data analysis. This is not a statistical task. It is informed by subject-matter knowledge and how the data were generated. For example, it matters if a variable that is important for the subject matter is absent in the collected data. Models that contain the omitted variable should be excluded from the universe of models. No such models could be realized not matter how well they correspond to subject-matter knowledge.

For that modeling universe, there can be a limitless number of i.i.d data realizations. For each realization of the dataset and each possible model there are parameter estimates. The estimate for any given parameter will usually vary over models unless the corresponding predictor is uncorrelated with all other predictors. And even for that same parameter and same model, estimates likely will vary over data realizations. 

Using this formulation, there is a multidimensional confidence region containing, with some high probability such as .95, \textit{all} of the parameter estimates over the full universe of models. The confidence region has the same number of dimensions as the number of parameters whose estimates are being sought. For a single parameter, the confidence region is a line representing a conventional confidence interval. If there are two parameters, the region is  a plane. If there are three such parameters, the confidence region is a solid such as an ellipsoid. With more than three dimensions one has an higher-dimensional solid such as a polytrope.

There is a duality between confidence regions and statistical tests (Rice, 2007: Section 9.3). A confidence region can be translated into one or more statistical tests and vice versa. This leads directly to the Bonferroi formulation based on the familywise error rate discussed in the previous section (Kuchibhotla et al., 2021: 8).  Asymptotically, the usual guarantee applies. 

As before, the Bonferroni correction is inherently conservative, although now for an additional reason. The $1-\alpha$ guarantee is over all models in the universe of models, not just for the selected model. One gains a form of insurance protection no matter what model the selection processes determines, but at a price. That price is arguably worth paying for interrupted time series models. Recall that no claims are made that a chosen model is in any sense the correct model or uniquely preferred. 

As discussed earlier, there are more powerful $1-\alpha$ guarantees that can work for the familywise error rate. The max-t approach is one. Because it relies on the bootstrap, dependence between estimates is properly incorporated automatically, unlike for the Bonferroni approach. Power is increased. And the resampling results can be used more widely than for statistical tests or confidence intervals. 

There are other approaches to post-model-selection inference that do not rest on simultaneous inference for the full model universe. They exploit information from a selected model and its selection procedure (Kouchibhotla et al., 2021: Section 3). These methods also can offer some gains in power but are less desirable for an interrupted time series analysis. There is no reason to treat any approximate model as special, especially when it is not likely to be uniquely superior. In addition, each new selection method requires its own derived, multiplicity correction, and the data analyst must adhere to the prescribed model selection method no matter what the results. Also, dependent data can present complications.

\section{Data} 

Over the past year, there has been growing pressure in the U.S. to ``do something'' about virtually unfettered access firearms. A range of policy interventions has been proposed building in part on past experiences with firearm policy in the U.S. and elsewhere (Cook, 2018; Carlson, 2020; Negin et al., 2021). California has been an important testbed. The interrupted time series analysis to follow considers whether sales of handguns were affected by a California assault weapon ban legislated in 1999 and implemented in the 2000. 

There are 8400 daily observations from 1996 through 2018.\footnote
{
The data provided had a bit more than 8400 daily observations. But for the earliest several months, the data were incomplete. Those months were removed. 
}
Reported for each day is the number of handgun background checks undertaken. In California, these checks are considered a good proxy for handgun sales. California has universal background checks such that, in addition to sales through federally licensed firearm dealers, private sales must be processed by a federally licensed firearm dealer; both kinds of transactions are subject to a background checks. The terms ``background checks" and ``handgun sales'' will be used interchangeably.\footnote
{
Parenthetically, there is no evidence that California's universal background checks reduce firearm mortality (Castillo-Carniglia et al., 2019)
} 

\begin{figure}
\begin{center}
\includegraphics[scale=.5]{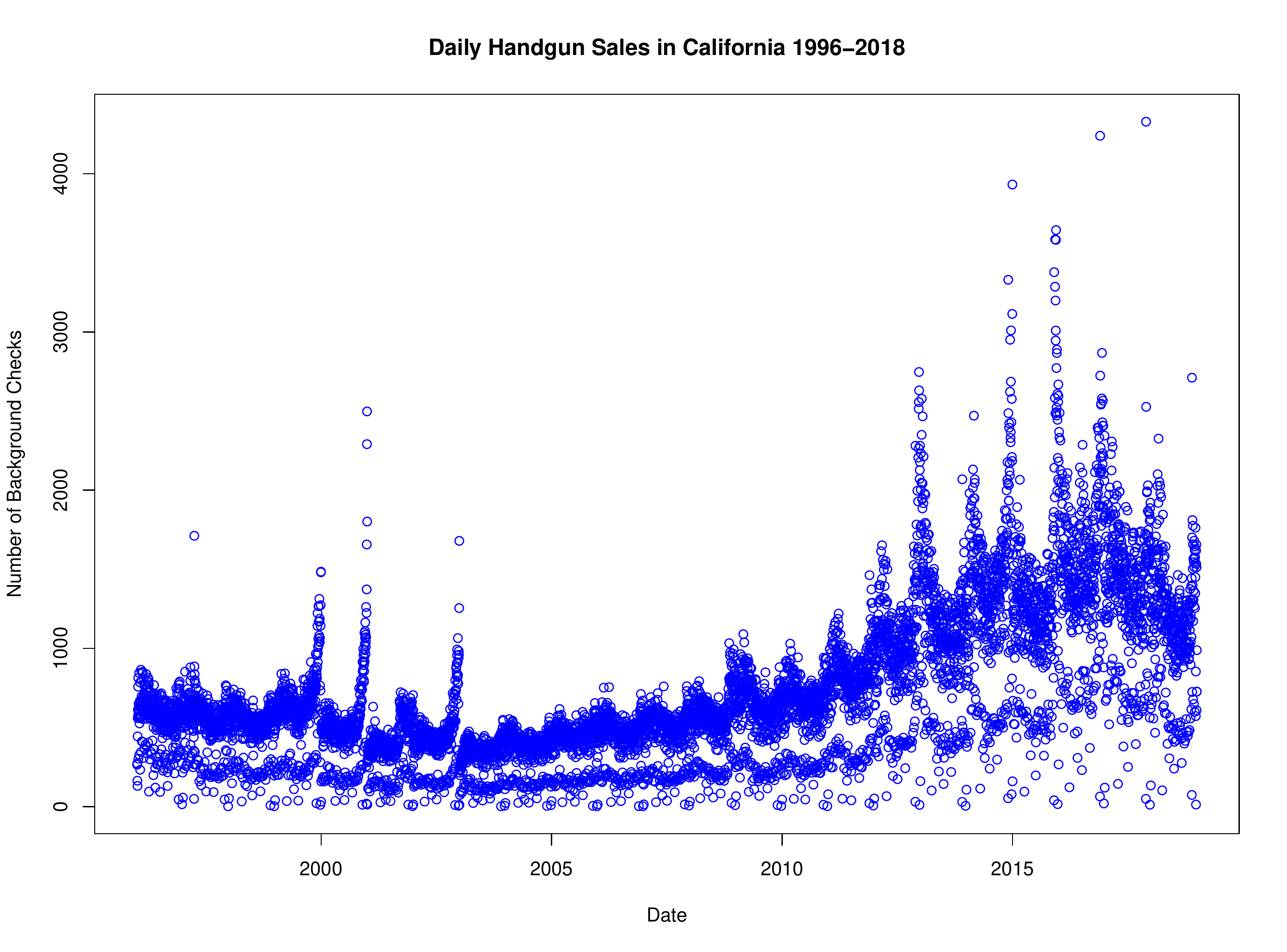}
\caption{The Number of Daily Handgun Background Checks in California Showing an Upward Trend and Large Peaks and Valleys (N=8400)}
\label{fig:seriessales}
\end{center}
\end{figure} 

Figure~\ref{fig:seriessales} is a time series plot of the sales data.\footnote
{
The figure and all to follow in this section were constructed after the interrupted time series model was specified and implemented. They were not used in model specification. They provide a context for the results reported in the Section 6. 
}
Its mean is 729 sales per day with a standard deviation is 467. Clearly, daily sales are substantial and growing with considerable day-to-day variation. There are also many dramatic peaks and valleys, and a curious gap toward the bottom of the plot. Distinct differences exist between days when many dealers are likely be closed (i.e., Sundays and holidays) and days when it is business as usual.  For the analysis that follows, it is unlikely that the gap makes a material difference because Sundays and holidays are included as an indicator variable in the interrupted times series model. 

\begin{figure}
\begin{center}
\includegraphics[scale=.50]{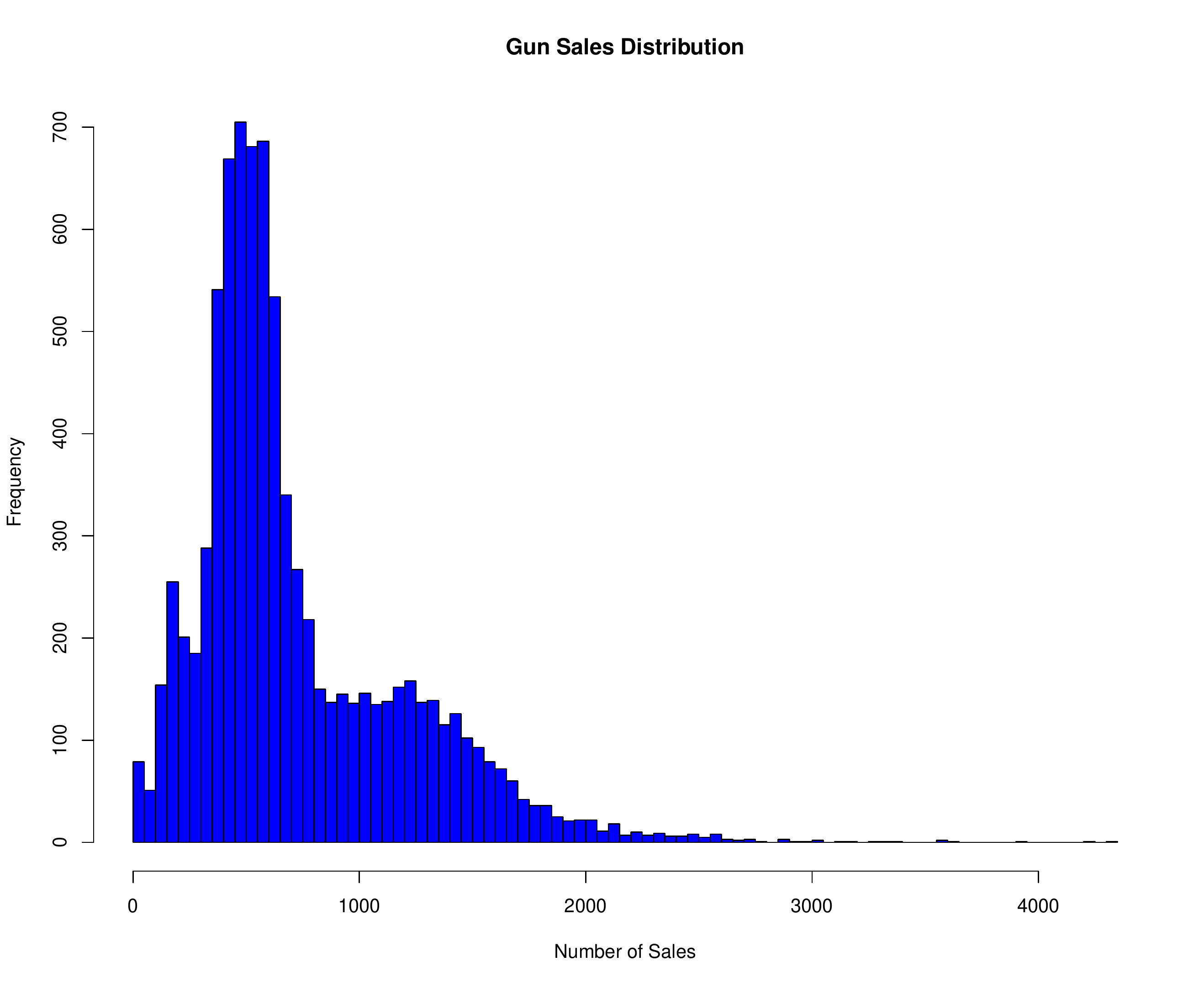}
\caption{Histogram of Daily Number of Handgun Background Checks (N=8400)}
\label{fig:sales}
\end{center}
\end{figure} 

The large peaks and valleys are revealed differently in Figure~\ref{fig:sales}. The histogram has a long right tail with values nearly an order of magnitude larger than most daily sales figures. There are also some days with less that 100 sales. The distribution poses difficult analysis challenges. 

Much as in the spirit of RCTs, the impact of the assault weapons ban is the single intervention of interest. Although within other design traditions one could certainly try to construct a general model of factors affecting handgun sales, for an interrupted time series analysis, other predictors treated as covariates are included primarily to adjust for possible confounding and increase precision.

\begin{figure}
\begin{center}
\includegraphics[scale=.50]{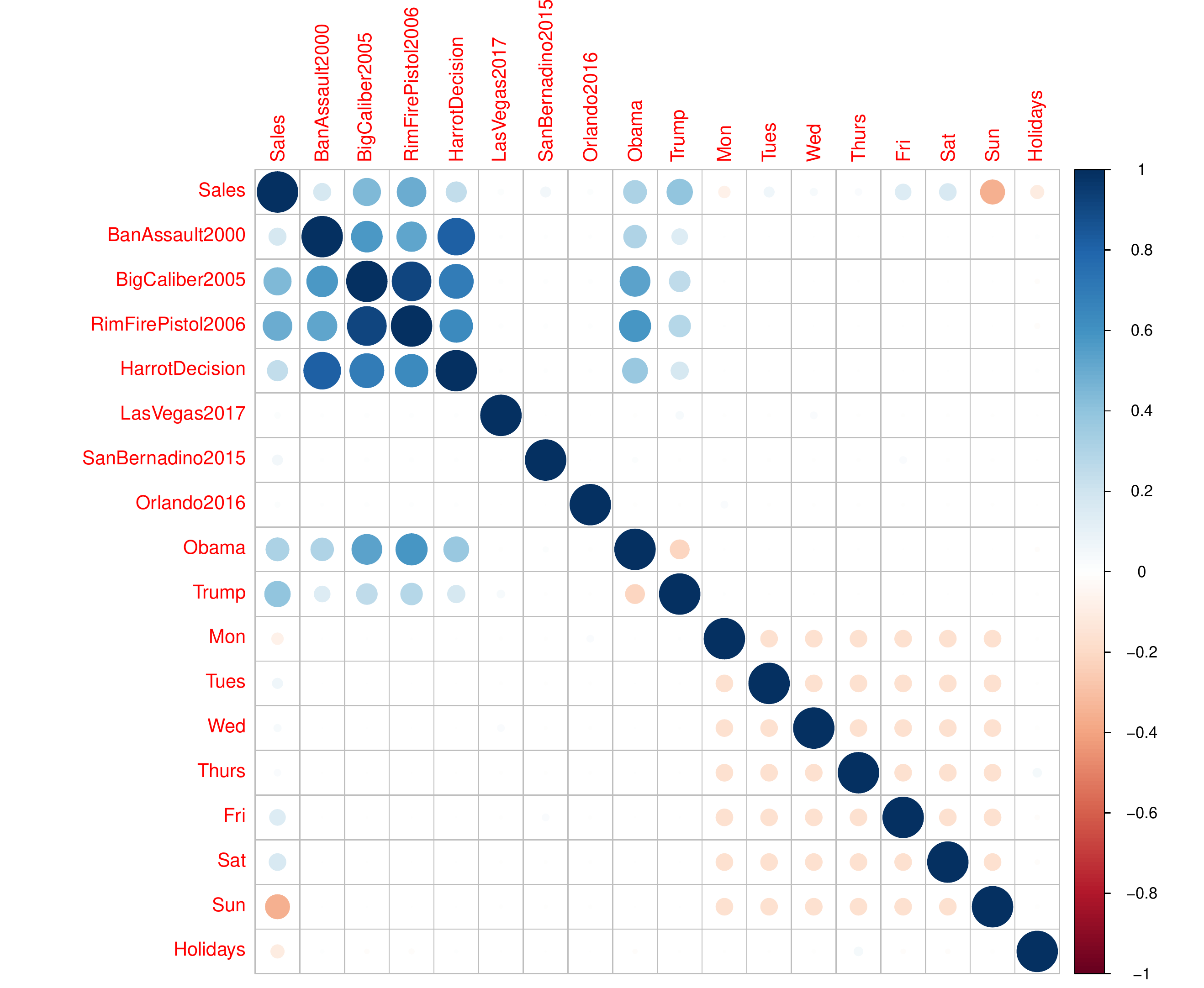}
\caption{Correlations Between Candidate Variables for the Interrupted Time Series Analysis}
\label{fig:corplot}
\end{center}
\end{figure} 

From that perspective, Figure~\ref{fig:corplot} shows the correlations between candidate covariates as well as their associations with daily handgun sales. In order from top to bottom on the left margin are the sales outcome variable, four legislative or judicial interventions that might directly affect firearm sales, three widely publicized mass shootings of the sort linked previously to firearm sales in California (Studdert et al., 2017; Webster, 2017)), the terms of President Obama and President Trump, also thought to be related to firearm sales (Depetris-Chauvin, 2015; Studdert et al., 2017), days of the week, and holidays. Although Figure~\ref{fig:corplot} might have influenced the model specification, it was on purpose constructed afterwards.

From 1996 through 2018, there were over 30 legislature initiatives or court decisions that might affect handgun sales. Of primarily interest, determined before looking at the data, was the only intervention that directly limited the sale of a very popular kind of firearm; Senate Bill 23 clarified and expanded the kinds assault weapons banned by the California Roberti-Roos Assault Weapons Control Act of 1989. SB 23 was passed by the state legislature in 1999 and became law on January 1st, 2000. Other intervention possibilities were the ban on .50 caliber rifles, the ban on rimfire pistols, and the Harrott court decision in June of 2001 that upheld the ban but required that banned assault weapons be identified by make and model, not just descriptive features such as barrel length.\footnote
{
Rimfire pistols and .50 caliber rifles were not in common use or in great demand. Rimfire pistols were small caliber (e.g. .22 caliber), but typically equipped with high capacity magazines, while .50 caliber rifles were designed for military use, and especially for snipers, who might require an effective range of a mile or more.  
}

One can see in Figure~\ref{fig:corplot} that all four legislative or legal interventions, coded as step functions, are positively correlated at modest to high levels. Had all four been included, multicollinearity likely would be a problem, especially for an interrupted time series analysis. 

 The San Bernardino mass shooting was the only such incident that occurred in California. San Bernardino lies 60 miles east of Los Angeles. Las Vegas is in Nevada and about 225 miles northeast of San Bernadino. Proximity favored San Bernardino. For this analysis, a mass shooting is represented by a pulse.

The presidential terms of Obama and Trump were separately coded as step functions, ``1'' while in office and ``0'' otherwise. Given the timing of Senate Bill 23, no dependency problems with the principal intervention were anticipated. Presidencies can raise concerns among some about 2nd amendment rights.  

Handgun sales vary by day of the week. But, including days of the week was also a potential source of multicollinearity. Because they are categories of the same nominal variable, there are necessarily negative correlations between them. As a potential solution, an indicator variable was constructed coded ``1'' for Sundays and holidays and otherwise ``0.'' 

The statistical analysis was undertaken in R with the procedure \textit{arimax} from the \textit{TSA} library. The five regressors $x_{1} \dots x_{5}$ were included as a linear combination, just as one would for a linear regression analysis.\footnote
{
More formally, $m_{t} = \omega_{0} + \omega_{1} x_{1} + \dots + \omega_{5} x_{5}$. No $\delta$ parameters were included (see Figure~\ref{fig:transfer}) because they would introduce additional lagged values of $y_{t}$. After the analysis was completed, exploration was undertaken with one or more $\delta$ parameters included. As expected, in almost every case, the fitting algorithm failed to converge. On the rare occasion when it did, there were warning messages.
} 
\textit{Every} model examined included the same five exogenous regressors: four covariates and one intervention. 
\begin{itemize}
\item $x_{1}$:
The Obama presidential terms as a step function, a covariate with a positive effect expected on handgun sales consistent with earlier studies and numerous anecdotes;
\item $x_{2}$:
The Trump presidential term as a step function, a covariate with a negative effect on handgun sales expected in contrast to the Obama presidency;
\item $x_{3}$:
Sundays and holidays as a pulses, a covariate with a negative effect on handgun sales expected because many firearm dealers are closed on such days;
\item $x_{4}$:
The San Bernardino mass shooting as a pulse, a covariate with an increase in handguns sales expected, consistent with earlier studies and numerous anecdotes; and
\item $x_{5}$:
The SB 23 assault weapons ban, the intervention of interest, included as a step function with a reduction in handgun sales expected.
\end{itemize}

A seasonal ARIMA formulation is commonly represented as ARIMA$(p,d,q) \times (P,D,Q)_{k}$, where $p$ and $P$ are the orders of AR terms, $d$ and $D$ are the orders of differencing, and $q$ and $Q$ are the orders of MA terms. The small letters specify the nonseasonal component. The capital letters specify the seasonal component with a seasonal lag of $k$. 

For the seasonal ARIMA component, a lag of 7 days seemed appropriate. Because the same days of the week were anticipated to affect handgun sales in similar ways, seasonality nonstationarity was likely. Hence, a first difference was included. The MA term was included to pick up short term dependence from random perturbations for the same days of the week. The seasonal specification was determined before the data analysis began and is a form commonly used for seasonal models. 
 
In contrast, the terms for the nonseasonal component were allowed to vary. The arguments $p, d,$ and $q$ each could take on a value of 0, 1, or 2 because any combinations of these values seemed reasonable a priori. For example, before looking at the data, there was anecdotal information that handgun sales had been increasing over time, which would ordinarily make differencing worth exploring. Short-term dependence is a universal concern in most time series data, which motivated a consideration of AR and MA terms.

This thinking led to 27 combinations of values yielding 27 models as the relevant model universe. Estimates for each model parameter for each model were computed, and the usual model diagnostics were evaluated (Box et al., 2016: Chapter 8): autocorrelation functions for the residuals, partial autocorrelation functions for the residuals, the augmented Dickey-Fuller t-statistic for stationarity (Box et al. 2016: Section10.1.2), and more.\footnote
{
Because of the very large number of observations, statistical tests were not helpful. Statistical power was excessive such that trivial departures from the null hypothesis were ``statistically significant.'' See Greenland et al. (2016) for a general discussion.
}

The preeminent goal was to find models for which the residuals were consistent with white noise. White noise residuals imply that all dependence in $y_{t}$ has been removed, including the impact of both gradual and abrupt confounders. Although one can construct a few (improbable) counter-examples, confounders that are not properly addressed by the mean function will usually be evident in the residual diagnostics. In this case, the residuals for each model were roughly consistent with white noise, implying that the mean function and seasonal component by themselves were a good start (i.e., no matter what form the nonseasonal ARIMA component took).  The model ultimately selected achieved a very close approximation of white noise residuals, had estimates of the intervention parameters with excellent precision, and offered plausible substantive interpretations.\footnote 
{
 Expanding the polynomials, the full noise model selected took the following form. $N_{t} = N_{t-1} + N_{t-7} - N_{t-8} + a_{t} - \theta_{1} a_{t-1} - \theta_{2} a_{t-7} + \theta_{3} a_{t-8}.$
 }

\section{Results}

\begin{figure}
\begin{center}
\includegraphics[scale=.55]{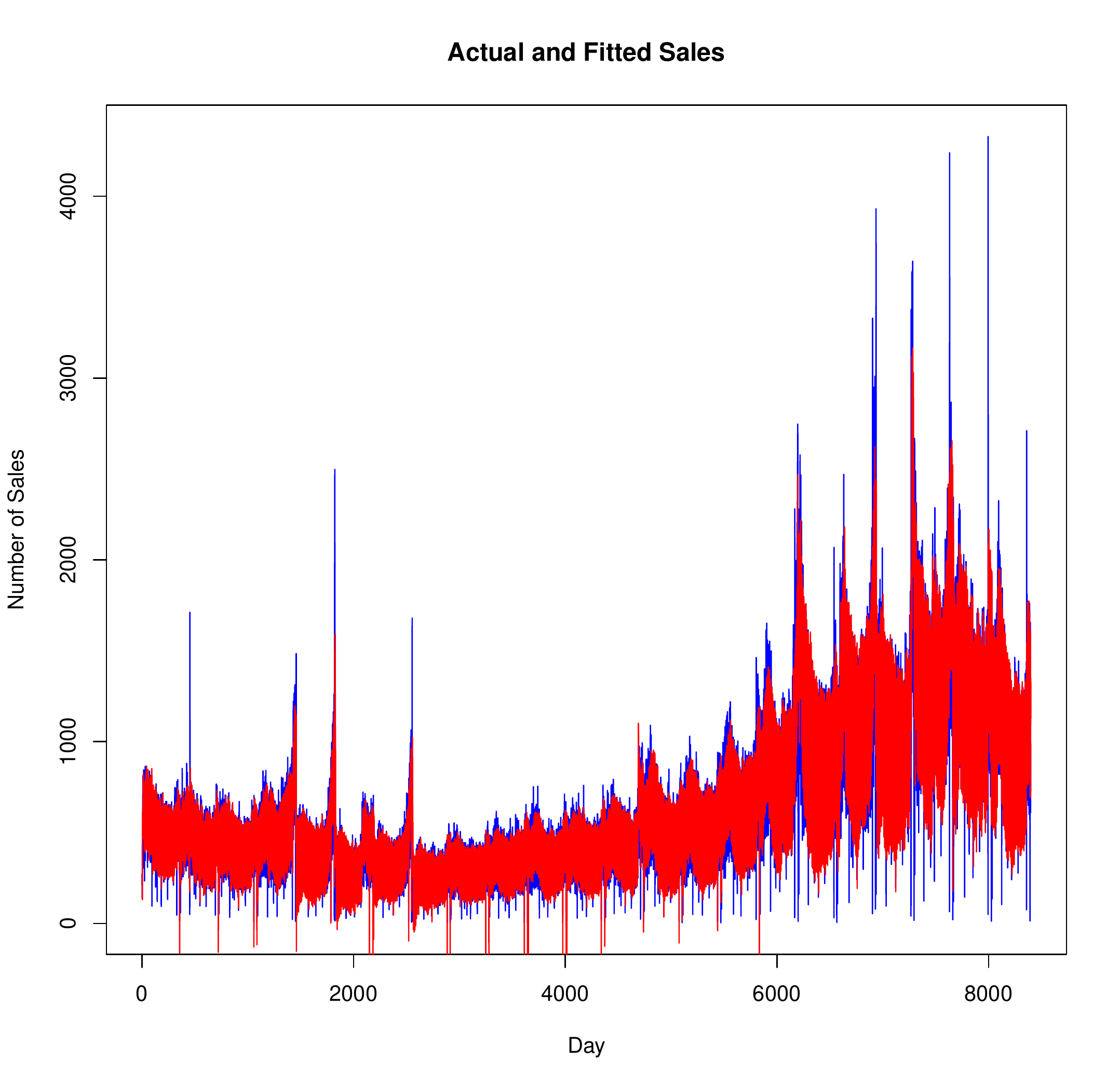}
\caption{Fitted Values Overlaid on Observed Background Checks with an Upward Trend and Many Peaks and Valleys Approximated Well (N=8400)}
\label{fig:fitted}
\end{center}
\end{figure}

As a sanity yardstick, Figure~\ref{fig:fitted} shows the correspondence between the actual number of daily background checks and the fitted number of daily background checks. Overall trends seem to be well captured along with many large peaks and valleys. There is no reason at this point to abort a more thorough model evaluation.

\begin{figure}
\begin{center}
\includegraphics[scale=.50]{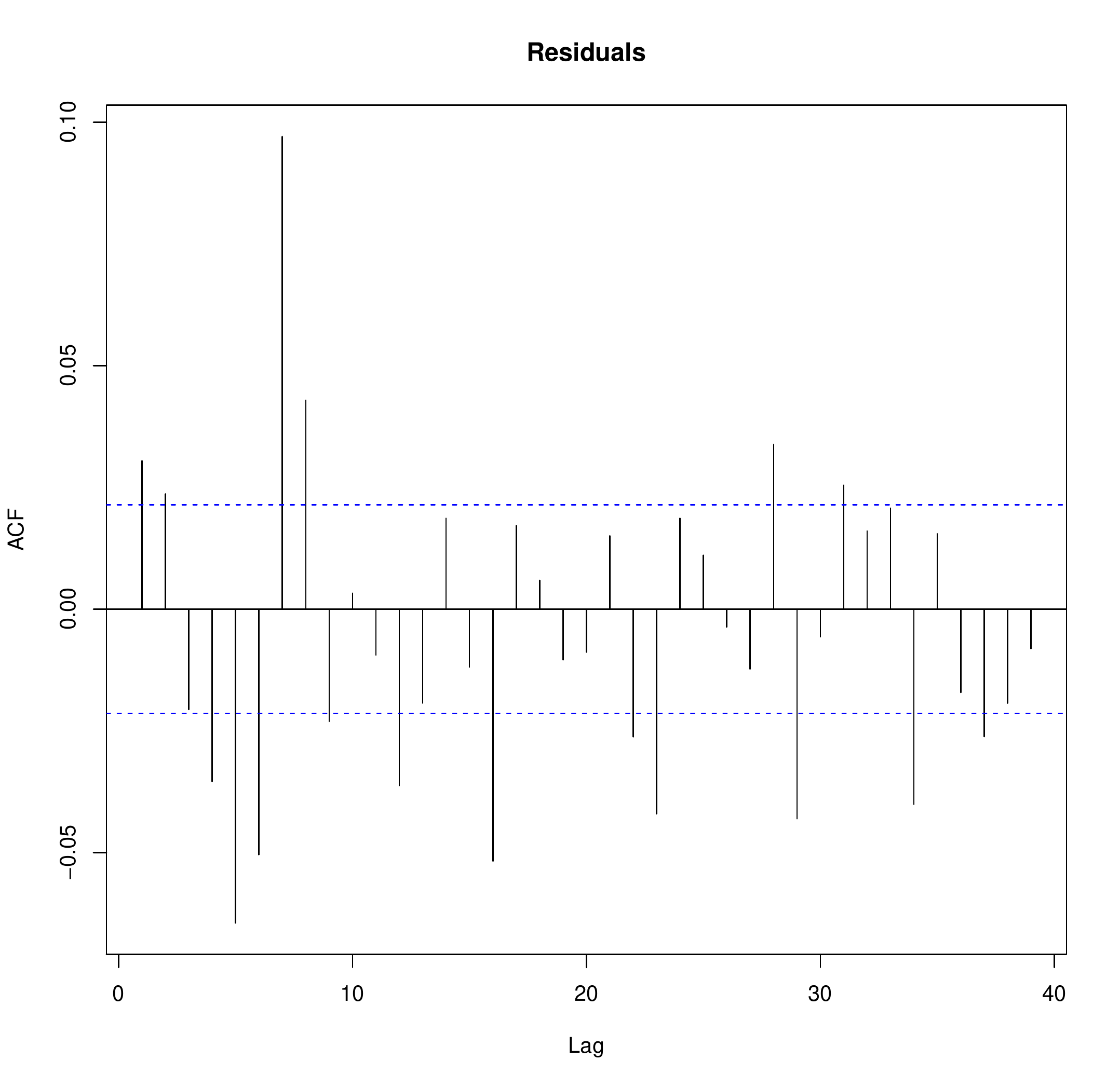}
\caption{Residual Autocorrelation Function  for the ``Best" Model}
\label{fig:acf}
\end{center}
\end{figure}

Figure~\ref{fig:acf} shows the results for the autocorrelation function of the residuals.
The largest autocorrelation is less than .10. Autocorrelations less than .20 rarely cause estimation problems. The horizontal dashed lines show plus or minus two standard errors. With 8400 observations, the error band is very narrow at autocorrelations of $\pm .025$. Clearly,  there is excessive power. In fact, the residuals look to be a good approximation of white noise.\footnote
{
One could respecify the noise model to remove the small spike at a lag 7 days. But that is likely to be an exercise in ``whack-a-mole.'' Another very small spike would likely appear at different lag. 
}

\begin{figure}
\begin{center}
\includegraphics[scale=.50]{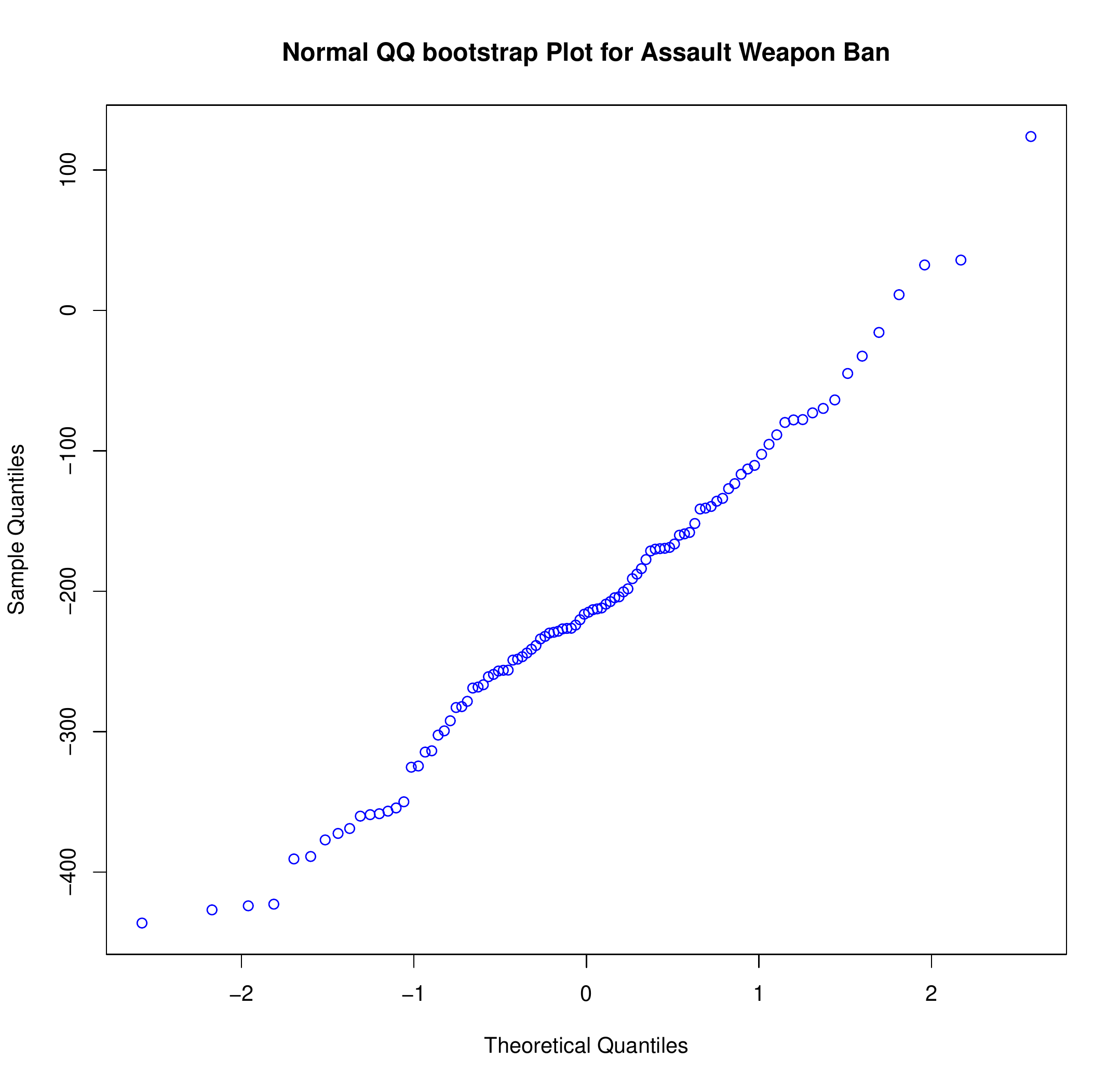}
\caption{Bootstrap Sampling Distribution for the Assault Weapons Ban Estimate Showing an Approximately Normal Sampling Distribution}
\label{fig:qq}
\end{center}
\end{figure}

Figure~\ref{fig:qq} shows a bootstrap sampling distribution for the assault weapons ban parameter estimate. The normal QQ plot is close to a straight line, consistent with an asymptotic normal distribution. The sample size apparently is sufficient, and the fitting algorithm properly converged for the selected model. 

Further encouraged, Table~\ref{tab:results} contains the estimates for all predictor coefficients and their associated t-statistics.\footnote
{
Coefficients for the noise model $N_{t}$ are rarely interpreted and are not shown in the table.
}
All but one of the coefficients have t-statistics well above the threshold needed for ``statistical significance'' at $\alpha = .05$, and larger t-statistics than needed for the recommended $\alpha = .005$ (Benjamin et al., 2018). Consistency between the estimates and the expected relationships adds credibility to the analysis. 

The assault weapons ban appears to reduce handgun sales by an extraordinary 634 a day, but uncertainty corrections need to be made for post-model-selection inference.  A conventional Bonferroni correction is provided for all statistical tests in Table~\ref{tab:results}, adjusting for the specification search over 27 models. The Bonferroni corrected t-statistic threshold is $\pm 2.87$, which is substantially larger than the unadjusted threshold of $\pm 1.96$. Nevertheless, all but one of the null hypotheses are still rejected. The effect of the assault weapons ban is still statistically significant at well beyond $\alpha = .0002$.\footnote
{
There is no need for simultaneous inference across all coefficients at once because the principal research question is the effect of the assault weapons ban. This is fully consistent with the intent of interrupted time series designs and analysis. The covariates are included to address confounding and prevision, not to inform subject matter questions, although in this case they are consistent with much earlier work.
}

\begin{table}[htp]
\caption{Coefficient Estimates and t-Statistics: Bonferroni Corrected t-Statistic Is $\pm 2.87$}
\begin{center}
\begin{tabular}{|l|c|c|}
\hline \hline
Interventions & Coefficient & t-statistic  \\
\hline
Obama & 462 & 7.11 \\
Trump & 42 & 0.46 \\
Sundays and Holidays & -714 & 3.15 \\
San Bernardino & 839 & 6.08 \\
Assault Weapon Ban & -634 & -8.57 \\
\hline \hline
\end{tabular}
\end{center}
\label{tab:results}
\end{table}

\begin{figure}
\begin{center}
\includegraphics[scale=.20]{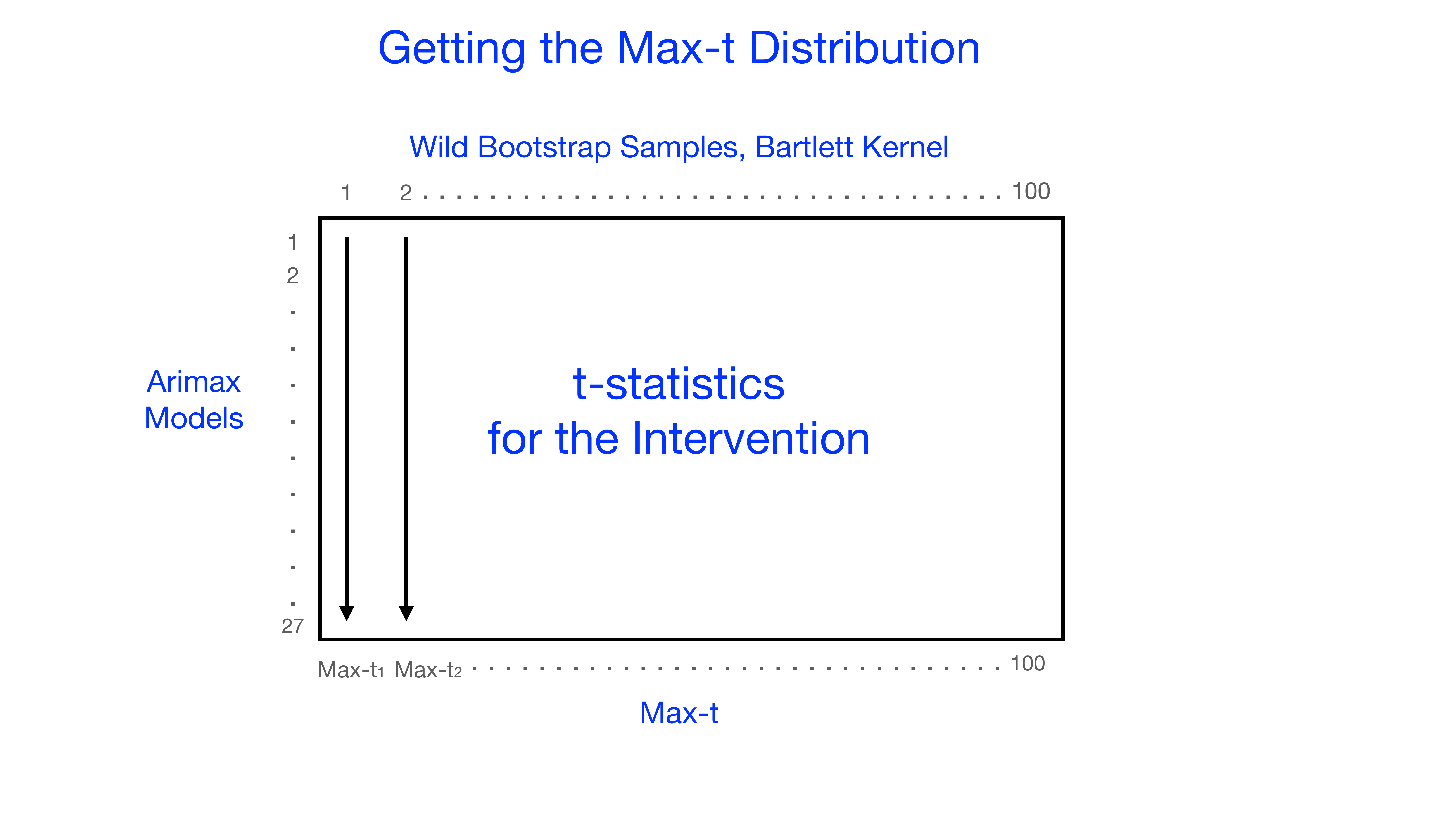}
\caption{Diagrammatic Explanation for Estimating the Max-t Distribution for the Assault Weapon Ban Coefficient}
\label{fig:wild}
\end{center}
\end{figure}

The Bonferroni correction is well known to be conservative. As introduced briefly earlier, we also applied a max-t correction. To account for dependence in the handgun sales data, we used a wild bootstrap with a Bartlett kernel to capture short term dependence of 7 days (Shao, 2010). Figure~\ref{fig:wild} is a diagrammatic rendering of the approach. For each of 100 bootstrap samples, all 27 models were fitted and the largest t-statistic for the 27 models stored, leading to 100 max-t values. The 100 values provided an estimate of the max-t sampling distribution from which appropriate thresholds were calculated by calibration using the .975 and .025 quantiles. In this case, the results were about the same as for the Bonferroni correction and are, therefore, not reported in Table~\ref{tab:results}. The likely explanation for no apparent improvement in power is the dependence in the handgun sales time series. Resulting instability created a long tailed coefficient distribution. Evidence is discussed shortly when Figure~\ref{fig:dist} is examined.

If the discussion of results ended at this point, the interrupted time series dataset looks to have been properly analyzed and has yielded credible results. The trial and error approach to model specification inflated statistical power, which the multiplicity corrections effectively addressed. In this case, true power was sufficient.  The assault weapon ban appears to cause a dramatic reduction in handgun sales.

\subsection{Complications}

One must be clear about what the multiplicity correction is doing. For each model from the universe of models, the estimand of primary interest is the coefficient for the assault weapon ban. However, just as for conventional linear regression, that coefficient's definition depends on how the model is specified (Wickens, 1995: Section4.1; Weisberg, 2014: Section4.1.3). Different models necessarily define the estimand differently because different adjustments for possible confounding are being undertaken. 

Whereas there is a single causal effect to be estimated for the assault weapons ban, there are 27 different estimands of that causal effect. Moreover, each model is explicitly an approximation of the linear, data generation process. One has 27 ``wrong" models with 27 disparate ``wrong'' estimands. 

Recent thinking in statistics (Buja et al., 2019a; 2019b) is consistent with such results. One has 27 different estimands, each of which is a``regression functional of interest.''\footnote
{
It can be helpful to think about each regression functional as a partial correlation coefficient in its original units. There are 27 different \textit{measures of the association} between the assault weapon's ban and daily handgun sales, depending on the model.  
}
Unfortunately,  there seems to be no way at the moment to get from a set of different causal estimands to credible causal inference about a single ``real world'' intervention.

\begin{figure}
\begin{center}
\includegraphics[scale=.50]{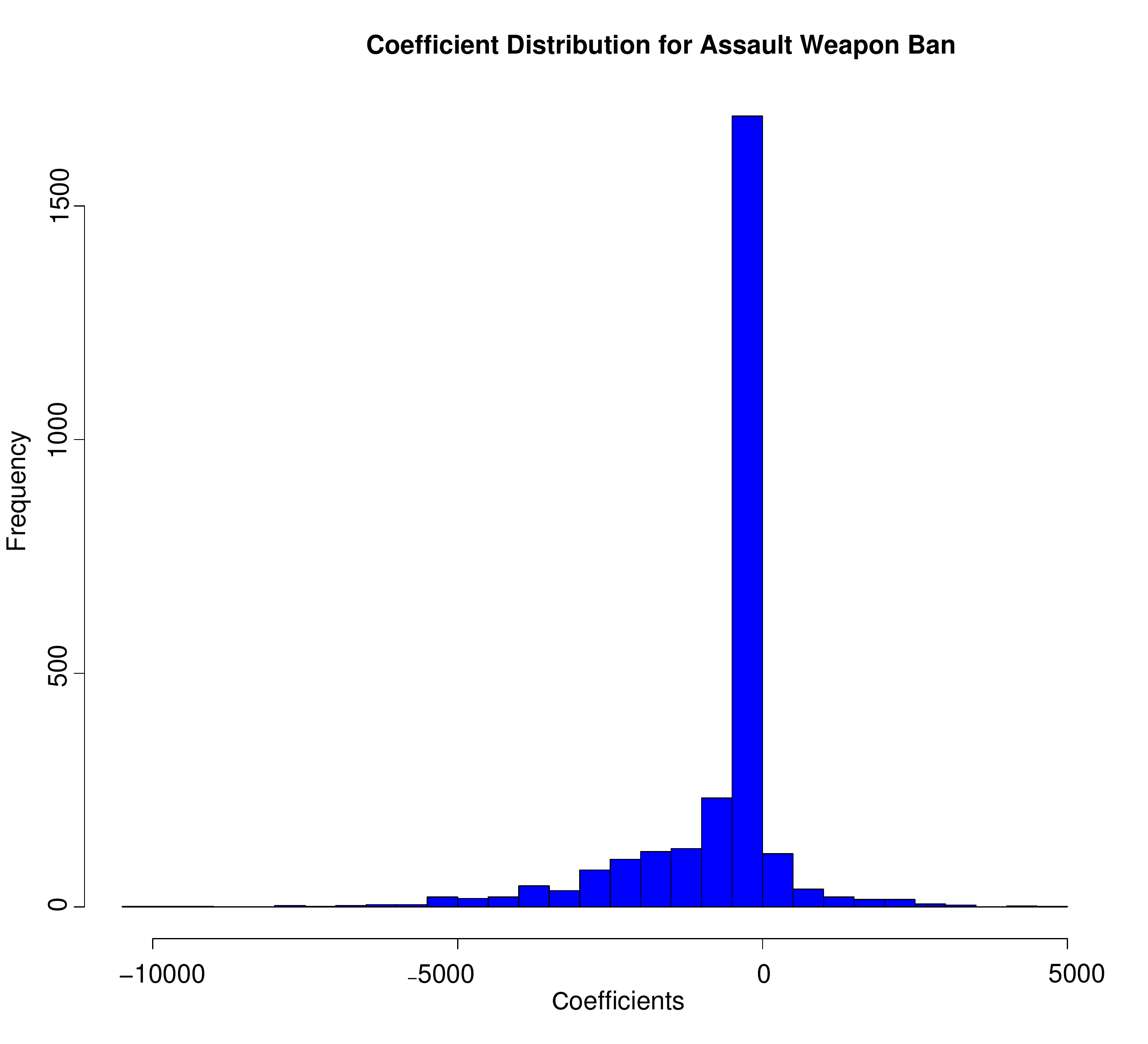}
\caption{Distribution of Assault Weapons Ban Coefficient Estimates Across the Universe of 27 Models: 94\% of the signs were negative.}
\label{fig:dist}
\end{center}
\end{figure}

Nevertheless, it may be possible to make a credible case at least for the sign of the regression functionals. If one has used a multiplicity correction based on resampling, such as max-t, one can examine the signs of the estimated functionals over models and samples. Ideally, positive or negative signs dominate. Figure~\ref{fig:dist} shows the coefficient distribution for the assault weapons ban estimates. 94\% of the signs are negative. Across the 27 models with 100 bootstrap samples each, negative associations dominate. The best bet is that the ban reduced sales. Other fallback positions will be discussed shortly.

Figure~\ref{fig:dist} also speaks to the earlier of concern about an absence of power gains for the max-t correction compared to the Bonferroni correction. A relatively large number of coefficients are to the left of  -1000. These are somewhat atypical and arguably implausible. Looking back at Figure~\ref{fig:seriessales} makes clear that from 1996 to 1999, before the assault weapon ban, there was only one day with 1000 sales or more. A decline of 1000 sales per day shortly after the ban would have consistently produced the curious statistical result of negative sales. Such difficulties can be caused by very unstable parameter estimates that, in turn, undermine meaningful power gains. Still, even for the questionable coefficient estimates, most of the signs are negative. The predominance of negative estimates is robust to model specification.

But for this particular application, unfortunately, there are major deficiencies in how well the assault weapons ban corresponds to a properly defined intervention. The ban applied to assault weapons and yet, background checks are a proxy for sales of handguns only. Although one can certainly speculate about how the two could be a causally connected, there is no consequential evidence available. 

The nature and timing of the intervention also is  unclear. Very public, heated discussions about an assault weapon ban began in 1999, months before the legislation was introduced. Media coverage was extensive. There were widely covered public hearings. Once AB 23 was passed, some time elapsed before it was signed by the governor. On January 1st of 2000, implementation began. In August of 2000, the original ``generic'' assault weapons ban passed in 1998 was affirmed in the Kasler v. Lockyer. The revisions under AB 23 were affirmed and clarified in Harrott v. County of Kings in June of 2001. Subsequently, a variety of gun-related legislation, such as a ban on high capacity magazines, was passed, also with considerable publicity. 

The nature and timing of the intervention has implications for the possible causal mechanisms operating. If reductions in sales stems from actions of gun dealers, who refused to make sales in violation of the assault weapons ban, then an intervention date of January 1st, 2000 might be justified. Yet, why were handguns sales affected? And what about the role of legal challenges and other legislation? If the reduction in sales stems from the actions of prospective handgun buyers, publicity surrounding may be important. In that context, an intervention date of January 1st 2000 seems arbitrary. 

After examining the results, the data were re-analyzed twice, in an exploratory fashion, using the selected model with the intervention date moved forward two months or back two months. For both, the estimated coefficient for the assault weapons ban became small and positive, much like the estimated effect of the Trump presidency. This evidence underscores the risks inherent in trying to determine the intervention's timing and content. In retrospect, such challenges probably are common for a wide variety social interventions responding to hot-button issues.

\section{Conclusions}

A credible analysis of data from an interrupted time series design requires that the intervention(s) be clearly defined and the appropriate locations in time determined. For legislative interventions, this can require substantial background information on the legislative history, related legislative initiatives, judicial decisions, and media coverage. It also helps if potential causal mechanisms linking an intervention to an outcome can be clearly articulated. These recommendations probably apply to a wide variety interventions well beyond gun policy. 

It is best to avoid data-driven model specification. When possible, relying on earlier research and credible theory is ideal. At a minimum, the mean function should be specified before looking at the data. Because several different noise functions likely will remove any remaining dependence, a relatively simple seasonal ARIMA formulation may well to suffice. After examining the results, model diagnostics should be employed and perhaps alternative models estimated. Statistical tests and confidence intervals no longer  have their usual validity guarantees and should be avoided. These explorations go to the credibility of the model originally selected. They are \textit{not} a means to replace the first model with a preferred alternative.  This is just a restatement of the usual confirmatory/exploratory distinction (Tukey, 1978). 

If inductive model selection cannot be avoided, a multiplicity correction and related diagnostics should be employed. Statistical tests and confidence intervals can then have valid performance guarantees, although some corrections will be conservative and others may be challenging to properly implement. The results must be interpreted with caution because of unsolved problems linking multiple estimands to a single causal effect. A more prudent approach is to interpret the estimated causal effects as only a form of association, controlling for other predictors in the model. One has regression functionals of interest, nothing more.

Supporting evidence may be available in the distribution of parameter estimates over bootstrap samples if a resampling approach such as max-t is used. Ideally, the vast majority of signs will be positive or negative. The message will be that across the universe of potential models, one can determine the likely sign of any causal effect; does the intervention make things better or worse? For many interventions, a conclusion of this sort could be very instructive.

More generally, no single investigation is likely to be definitive. Credible evidence depends on a collection of well-executed studies. All research needs to build on earlier results. For the analysis of data from an interrupted time series design, working collaboratively within a scientific community is a way to help avoid post-model-selection problems. Models for later studies are determined by the results from earlier studies.

\section*{References}
\begin{description}
\item
Bachoc, F., Preinerstorfer, D., and Steinberger, L. (2020). ``Uniformly Valid Confidence Intervals Post-Model-Selection. \textit{The Annals of Statistics} 48(1):440 -- 463.
\item
Barnett, V., (1999) \textit{Comparative Statistical Inference}, third edition, Wiley.
\item
 Benjamin, D. J., Berger, J. O., Johannesson, M., Nosek, B. A., Wagenmakers, E. J., Berk, R.A., ...
(2018). ``Redefine Statistical Significance.'' \textit{Nature Human Behavior} https://doi.org/10.1038/s41562-017-0189-z 32
\item
Benjamini, Y., Hochberg, Y. (1995) ``Controlling the False Discovery Rate: A Practical and Powerful Approach to Multiple Testing.'' \textit{Journal of the Royal Statistical Society, Series B} 57 (1) 289--300.
\item
Benjamini, Y., Yekutieli, D, (2001). ``The Control of the False Discovery Rate in Multiple Testing Under Dependency'' (PDF). \textit{Annals of Statistics} 29 (4): 1165?1188
\item
Berk, R, A., Brown, L.,  Zhao, L. (2009). ``Statistical Inference After Model Selection.'' \textit{Journal Quantitative Criminology} 26 (2): 217 -- 236.
\item
Berk, R.A., Brown, L., Buja, A., George, E., Zhang, K., Zhao, L. (2013) ``Valid Post-Selection Inference,'' \textit{Annals of Statistics} 41(2), 2013.
\item
Box, G.E.P., and Tiao, G.C. (1975) ``Intervention Analysis with Applications to Economic and Environmental Problems.'' \textit{Journal of the American Statistical Association} 70(349): 70-79.
\item
Box,  G.E.P., Jenkins, G.W., Reinsel, G.C., and Ljung, G.M. (2016) \textit{Time Series Analysis: Forecasting and Control} Wiley.
\item
Bretz, F., Hothorn, T., and Westfall, P. (2011) \textit{Multiple Comparisons Using R}. CRC Press.
\item
Buja, A., Berk, R., Brown, L., George, E., Pitkin, E., Traskin, M., Zhan, K., and Zhao, L. (2019a). ``Models as Approximations -- Part I: A Conspiracy of Nonlinearity and Random Regressors Against Classical Inference in Regression.'' \textit{Statistical Science} 34(4): 523 -- 544.
\item
Buja, A., Berk, R., Brown, L., George, E., Arun Kumar Kuchibhotla, 
and Zhao, L. (2019b). ``Models as Approximations -- Part II: A General 
Theory of Model-Robust Regression.'' \textit{Statistical Science} 34(4):
545 -- 565.
\item
Castillo-Carniglia, A.,Kagawa, R.M.C., Cerd\'{a}, M., Crifasi, N., Vernick, J.S., Webster, D.W., and Wintemute, G.J. (2019) ``California's Comprehensive Background Check and Misdemeanor Violence Prohibition Policies and Firearm
Mortality.'' \textit{Annals of Epidemiology} 30: 50 -- 56.
\item
Campbell, D.T., (1969) ``Reforms as Experiments.'' \textit{The American Psychologist} 24(4): 409 -- 429.
\item
Campbell, D.T. and Stanley, J. (1963) \textit{Experiental and Quasi-Experimental Designs for Resaerch}. Rand McNally.
\item
Campbell, D.T., and Ross, H.L. (1968) ``The Connecticut Crackdown on Speeding: Time-Series Data in Quasi-Experimental Analysis.'' \textit{Law and Society Review} 3(1): 33
\item
Carlson, J. (2020) ``Gun Studies and the Politics of Evidence.''  \textit{Annual Review of Criminology} 16: 183 -- 202.
    \item
Cook, P.J. (2018) ``Gun Markets.'' \textit{Annual Review of Criminology} 1: 359 -- 377.
\item
Cryer, J.D., and Chan, K-S (2008) \textit{Time Series Analysis with Applications in R}, Springer.
\item
Depetris-Chauvin, E. (2015) ``Fear of Obama: An Empirical Study of the Demand for Guns and U.S. 2008 Presidential Election.'' \textit{Journal of Public Economics} 130: 55 -- 79.
\item
Efron, B., Tibshirani, R.J. (1993) \textit{Introduction to the Bootstrap}. Chapman \& Hall.
\item
Gottman, J.M. and Glass, G.V. (1978) ``Analysis of Interrupted Time Series Experiments.'' in T.R. Kratochwill (ed.), \textit{Single Subject Research},  Academic Press.
\item
Granger, C.W. (1980) \textit{Forecasting on Business and Econimics}. Emerald Group Publishing.
\item
Greenland, S., Senn, S.J., Rothman, K.J., Carlin J.B., Poole, C., Goodman, S.N., and Douglas G. Altman, D.G. (2106) ``Statistical Tests, P Values, Confidence Intervals, and Power: A Guide to Misinterpretations.'' \textit{European Journal of Epidemiology} 31: 337-- 350.
\item
Hamilton, J.D. (1994) \textit{Time Series Analysis} Princeton University Press.
\item
Hendry, D.F. (1995) \textit{Dynamic Econometrics} Oxford University Press.
\item
Holm, S. (1979) ``A Simple Sequentially Rejective Multiple Test Procedure.'' \textit{Scandinavian Journal of Statistics} 6: 65--70.
\item
Holmes, M.D., Daudistel, H.C. and Taggert, W.A.(1992) ``Plea Bargaining Policy and State District Court Caseloads: An Interrupted Time Series Analysis.'' \textit{Law \& Society Review} 26(1): 133 -- 160.
\item
Ioannidis, J. P. A. (2005). ``Why Most Published Research Findings Are False.'' \textit{PLOS Medicine} 2 (8): e124.
\item
Ioannidis, J. P. A. (2012) ``Why Science Is Not Necessarily Self-Correcting.'' \textit{Perspectives on Psychological Science} 7 (6): 645--654.
\item
Kolaczyk, E.E., (2009) \textit{Statistical Analysis of Network Data}, Springer.
\item
Kuchibhotla, A.K., Kolassa, J.E., and Kuffner, T.A. (2021) ``Post-
Selection Inference.'' \textit{Annual Review of Statistics and Applications}, forthcoming.
\item
Lee, J., Sun, D., Sun, Y. and Taylor, J. (2016) ``Exact post-selection inference, with application to the lasso.'' \textit{Annals of Statistics} 44(3): 907 -- 927.
\item
Leeb, H., and P\"{o}tscher, B.M. (2005) ``Model Selection and Inference: Facts and Fiction,'' \textit{Econometric Theory} 21: 21--59.
\item
Leeb, H., and P\"{o}tscher, B.M.  (2006) ``Can one Estimate the Conditional Distribution of Post-Model-Selection Estimators?'' \textit{The Annals of Statistics} 34(5): 2554--2591.
\item
Leeb, H., P\"{o}tscher, B.M  (2008) ``Model Selection,'' in T.G. Anderson, R.A. Davis, J.-P. Kreib, and T. Mikosch (eds.), \textit{The Handbook of Financial Time Series}, New York, Springer: 785--821.
\item
Levine, P.B., and McKnight, R. (2017) ``Firearms and Accidental Deaths: Evidence from the Aftermath of the Sandy Hook School Shooting,'' \textit{Science} 358(6368): 1324--1328.
\item
Liu, G., and Wiebe, D.J. (2019) ``A Time-Series Analysis of Firearm Purchasing After Mass Shooting Events in the United States.'' \textit{Journal of the American Medical Association Netw Open} 2(4): e191736.
\item
Maxwell, D.E., Delaney, H.D., and Kelly, K. (2017) \textit{Designing Experiments and Analyzing Data: A Model Comparison Perspective} Third Edition, Routledge. 
\item
McDowall, D., McCleary, R., Meidinger, E.E., and Hay, R.A. (1980) \textit{Interrupted Time Series Analysis}. Sage Publications.
\item
McDowall, D., McCleary, R. and Bartos, B.J. (2019) \textit{Interrupted Time Series Analysis}. Oxford University Press.
\item
Negin, J., Alpers, P., Nassar, N.,and  Hemenway D. (2021) ``Australian Firearm Regulation at 25 -- Success, Ongoing Challenges, and Lessons for the World.'' \textit{New England Journal of Medicine} 384(17): 1581 -- 1582.
\item
O'Carroll, P.W, Loftin, C., Waller, J.B.,McDowall, D., Bukoff, A., Scott, R.O., Mercy, J.A.,  Wiersema, B. (1991) ``Preventing Homicide: An Evaluation of the Efficacy of a Detroit Gun Ordinance.'' \textit{American Journal of Public Health} 81: 576 -- 581.
\item
Rice, J.A. (1995) \textit{Mathematical Statistics and Data Analysis}, second edition, Duxbury Press.
\item
Romano, J.P., Wolf, M. (2005) ``Exact and Approximate Stepdown Methods for Multiple Hypithesis Testing.'' \textit{Journal of the American Statistical Association} 100 (5): 94--108.
\item
Romano, J.P., Wolf, M. (2017) ``Multipler Testing of One-Sided Hypotheses: Combining Bonferroni and the Bootstrap.'' Working Paper No.254, Department of Economics, University of Zurich.
\item
Schooler, J. W. (2014). ``Metascience Could Rescue the Replication Crisis." \textit{Nature} 515 (7525): 9. 
\item
Shao, X., (2010) ``The Dependent Wild Bootstrap.'' \textit{Journal of the American Statistical Association} 105 (489): 218 -- 235.
\item
Shiraishi, T-A., Sugiura, H., Matsuda, S-I. (2019) \textit{Pairwise Multiple comparisons: Theory and Computations} Springer
\item
Studdert, D.M., YZhang, Y., Rodden, J.A., Hyndman, R.J.,Wintemute, G.J. (2017) ``Handgun Acquisitions in California After Two Mass Shootings.'' \textit{Annals of Internal Medicine} 166(10): 698 -- 706
\item
Tibshirani, R. J., Taylor, J., Lockhart, R., and Tibshirani, R. (2016) ``Exact Post-Selection Inference for Sequential Regression Procedures.'' \textit{Journal of the American Statistical Association} 111 (514): 600 -- 620.
\item
Tukey, J.W. (1978) ``We Need Both Exploratory and Confirmatory.'' \textit{The American Statistician} 34 (1): 23- -25. 
\item
Vuji\'{c}, S., Commandeur, J.J.F., and Koopman, S.J. (2016) ``Intervention Time Series Analysis of Crime Rates: The Case of Sentence Reform in Virginia.'' \textit{Economic Modeling} 57: 311 -- 323.
\item
Webster, D.W. (2017) ``The True Effect of Mass Shootings on Americans.'' \textit{Annals of Internal Medicine} 166 (10): 749 -- 750.
\item
Webster, D.W., Venick, J.S., and Bulzacchelli, M.T. (2006) ``Effects of a Gun Dealer's Change in Sales Practices on the Supply of Guns to Criminals.'' \textit{Journal of Urban Health} 83: 778 -- 787.
\item
Weisberg, S. (2014) \textit{Applied Linear Regression} Wiley.
\item
Wickens, T.D. (1995) \textit{The Geometry of Multivariate Statistics} Lawrence Erlbaum.
\end{description}

\end{document}